\documentclass[aip,jcp,reprint]{revtex4-2}

\usepackage{framed}
\usepackage{graphicx}

\usepackage{color}

\definecolor{shadecolor}{gray}{0.96}

\usepackage{graphicx}
\usepackage{color}
\usepackage{amsbsy}
\expandafter\let\csname equation*\endcsname\relax
\expandafter\let\csname endequation*\endcsname\relax 
\usepackage{amsmath}

\newcommand{\dbar}{\mathchar'26\mkern-12mu d}

\def\l{\lambda}

\def\beq{\begin{equation}}
\def\ee{\end{equation}}
\def\bi{\begin {itemize}}
\def\ei{\end{itemize}}

\def\nb{n_{\beta}}

\def\lsim
{\protect \raisebox{-0.75ex}[-1.5ex]{$\;\stackrel{<}{\sim}\;$}}
\def\gsim
{\protect \raisebox{-0.75ex}[-1.5ex]{$\;\stackrel{>}{\sim}\;$}}
\def\lsimeq
{\protect \raisebox{-0.75ex}[-1.5ex]{$\;\stackrel{<}{\simeq}\;$}}
\def\gsimeq
{\protect \raisebox{-0.75ex}[-1.5ex]{$\;\stackrel{>}{\simeq}\;$}}

\def\l{\lambda}

\def\St{S_{\rm tot}(\tau)}

\def\c{{\cal C}}
\def\N{\mathcal{N}}

\def\tot{_{\rm tot}}
\def\cgr{_{\rm cgr}}
\def\wtd{_{\textrm{WTD}}}
\def\emc{_{\textrm{EMC}}}

\def\res{_{\textrm{res}}}
\def\mech{^{\textrm{mech}}}
\def\chem{^{\textrm{chem}}}

\def\Ic{{\mathcal{I}}}
\def\Jc{{\mathcal{J}}}
\def\Kc{{\mathcal{K}}}
\def\Oc{{\mathcal{O}}}
\def\xb{{\mathbf{x}}}

\def\Fb{{\mathbf{F}}}
\def\jb{{\mathbf{j}}}
\def\zb{{\boldsymbol{\zeta}}}
\def\nb{{\mathbf{n}}}
\def\nablab{{\boldsymbol{\nabla}}}

\def\DKL{\mathsf{D}}
\def\dkl{_{\textrm{KLD}}}

\def\kbT{\mathsf{k_BT}}
\def\kb{\mathsf{k_B}}
\def\kbs{\mathsf{k_B}/\textrm{s}}
\def\TT{\mathsf{T}}
\def\mum{\mu\textrm{m}}

\def\nub{{\boldsymbol \nu}}

\def\beq{\begin{equation}}
\def\ee{\end{equation}}

\def\bi{\begin {itemize}}
\def\ei{\end{itemize}}

\def\nn{\nonumber}

\begin{document}

\title{Universal bounds on entropy production from fluctuating coarse-grained trajectories
}
\author{
 Udo Seifert 
}

\affiliation{
{II.} Institut f\"ur Theoretische Physik, Universit\"at Stuttgart,
  70550 Stuttgart, Germany}
\begin{abstract}
Entropy production is arguably the most universally applicable measure of non-equilibrium behavior, particularly for systems coupled to a heat bath. This setting encompasses driven soft matter as well as biomolecular, biochemical, and biophysical systems. Despite its central role, direct measurements of entropy production remain challenging—especially in small systems dominated by fluctuations. The main difficulty arises because not all degrees of freedom contributing to entropy production are experimentally accessible. A key question, therefore, is how to infer entropy production from coarse-grained observations, such as time series of experimentally measurable variables.
Over the past decade, stochastic thermodynamics has provided several inequalities that yield model-free lower bounds on entropy production from such coarse-grained data. The major approaches rely on observations of coarse-grained states, fluctuating currents or ticks, correlation functions of coarse-grained observables, and waiting-time distributions between so-called Markovian events, which correspond to transitions between mesoscopic states.
Here, we systematically review these techniques valid under the sole assumption of a Markovian, i.e., memoryless, dynamics on an underlying, not necessarily observable, network of states or following a possibly high-dimensional Langevin equation.  We discuss in detail the large class of non-equilibrium steady states and highlight extensions of these methods to time-dependent and relaxing systems. While our focus is on mean entropy production, we also summarize recent progress in quantifying entropy production along individual coarse-grained trajectories.
\end{abstract}

\maketitle

\def\lsim
{\protect \raisebox{-0.75ex}[-1.5ex]{$\;\stackrel{<}{\sim}\;$}}

\def\gsim
{\protect \raisebox{-0.75ex}[-1.5ex]{$\;\stackrel{>}{\sim}\;$}}

\def\lsimeq
{\protect \raisebox{-0.75ex}[-1.5ex]{$\;\stackrel{<}{\simeq}\;$}}

\def\gsimeq
{\protect \raisebox{-0.75ex}[-1.5ex]{$\;\stackrel{>}{\simeq}\;$}}

\def\k{^{\rm k}}
\def\c{^{\rm c}}

\def\dek{(\Delta \Ek)^2}

\def\ph{p^{\rm hist}}

\def\pb{{\bf p}}
\def\qb{{\bf q}}
\def\xb{\mathbf{x}}
\def\ab{\mathbf{a}}

\def\ijI{{ij \in \Ic\Jc}}

\def\app{_{\textrm{app}}}
\def\tur{_{\textrm{TUR}}}
\def\osc{_{\textrm{osc}}}
\def\var{{\textrm{var}}}

\def\xib{{\boldsymbol \xi}}
\def\H{{\cal H}}


\def\call{}
\def\W{{\cal  W}}
\def\E{{\cal E}}
\def\Q{{\cal  Q}}
\def\St{{\cal  S}_{\rm tot}}
\def\Ss{{\cal  S}}
\def\Sm{{\cal  S}_{\rm b}}
\def\st{s_{\rm tot}}

%
%
%
%

\section {Introduction}

The second law of classical thermodynamics states that the entropy of the ``universe'' cannot decrease in any process. In many physical settings, this universe consists of a system of interest in  contact with a heat bath. The total entropy change then is the one of the system and that of the bath where the latter is related to the energy dissipated into it or taken up from it.  
This scenario is common in soft matter -- such as polymers, colloids, membranes --, in single-molecule experiments, in bio-molecular processes occurring in living cells, and in many active systems. All these systems are embedded in an aqueous solution that serves as a heat bath of well-defined temperature. Additionally, they may be driven by mechanical forces, such as those applied via optically trapped beads, and by chemical reactions like ATP hydrolysis.

Stochastic thermodynamics \cite{seki10,jarz11,seif12,vdb15,peli21,shir23,seif25}
provides a framework for the analysis of such systems that builds on a generalization of concepts from classical thermodynamics by getting rid off certain of its constraining implicit assumptions while still recovering the classical theory in the appropriate limits. It considers systems with arbitrary initial and final distribution, rather than the initial and final equilibrium states of the classical theory, and identifies thermodynamic quantities like work, heat and entropy production on the level of individual fluctuating trajectories, rather than  on the ensemble. A crucial assumption, however, is that the dynamics is effectively Markovian, i.e., memory-less. This key requirement is less constraining as it might appear on first sight since
it suffices to assume that this Markovian dynamics applies on a deeper underlying level of description.
Conceptually, this framework is thus consistent even if this level is not  experimentally accessible or resolvable. In such a situation, which is the common one for many -- if not most -- systems mentioned above, one then has to ask whether and how observations on a more coarse-grained higher level can reveal information about the system, in particular in the present context, about its physical entropy production.

The topic of this review is a survey of recent advances in thermodynamic inference that yield  lower bounds on entropy production based on measurements
of  coarse-grained observables. These measurements can then be used as input in a variety of theoretical results
that all lead to a lower bound on entropy production without assuming a specific model for the system under study. Such lower bounds are here called ``universal'' or ``model-free'' since they are applicable to any system that complies, on some deeper unobservable level,  with the central 
assumption of stochastic thermodynamics. Moreover, in order to keep the scope well-defined and bounded, we will focus on non-invasive techniques that do not require to perturb the system beyond the originally imposed driving. Finally, we will stay in the classical realm leaving out genuine open quantum systems.

We begin with systems in a non-equilibrium steady state (NESS)
for which we discuss operationally accessible bounds on the mean rate of entropy production using coarse-grained trajectory data.  Next, we show how to identify entropy production at the level of individual coarse-grained trajectories.
These approaches are then extended to encompass relaxation dynamics and time-dependent driving. We conclude with two sections outlining  experimental challenges and directions for further theoretical development.

Complementary reviews and perspectives include those  
on irreversibility in general \cite{luce25} and in living systems \cite{gnes18},  on the concept of thermodynamic inference \cite{seif19}, on the thermodynamic uncertainty relation (TUR) \cite{horo20}, on physical bioenergetics  \cite{yang21,brow24}, on irreversibility in active matter \cite{obyr22}, on signatures of irreversibility on the coarse-grained level \cite{dieb25}, and on macroscopic stochastic thermodynamics \cite{fala25}.

\section{Entropy production in a NESS}
%

A central assumption of stochastic thermdynamics is 
that, through an initial coarse-graining, the underlying microscopic degrees of freedom can be partitioned into meso-states $\{i,j,k\}$ (or $\{\bf x\}$, in the continuous case) such that the resulting dynamics  on this reduced state space becomes Markovian, i.e., memory-less.

For discrete meso-states as outlined in Box 1, this assumption leads to
the concept of transition rates $k_{ij}$ between meso-states $i$ and $j$. In a NESS,
this non-equilibrium system acquires a stationary probability $p_i$ with a mean rate of entropy production 
\cite{seif12,vdb15,peli21,shir23,seif25}
\beq\sigma = \sum_{ij}p_ik_{ij}
\ln(p_ik_{ij}/p_jk_{ji})\geq 0,
\label{eq:sigma}
\ee where we set Boltzmann's constant $\kb=1$ throughout.
This rate comprises the entropy production in the heat bath and in the chemical reservoirs since, in the steady state, the entropy of the system does not change in the mean.

The second class of models with a consistent stochastic thermodynamics
are systems that,
 on an intermediate level, can be described by continuous degrees of freedom $\xb$ like the position of a colloidal particle or the end-to-end vector of a biopolymer (and, for  a many-particle system, the collection of such quantities). The position $\xb(t)=\xb^t$ follows an overdamped Langevin equation
\beq\dot\xb(t) = \beta D \Fb(\xb^t,\l^t) + \zb(t) .
\ee The effective force $\Fb(\xb,\l)$ may be time-dependent through an external control parameter $\l(t)=\l^t$. The strength of the Gaussian white noise $\langle \zb(t)\otimes\zb(t')\rangle = 2 D \boldsymbol{1}\delta(t-t')$,
which arises from the environment at inverse temperature $\beta=1/\kbT$, is given by the bare diffusion constant $D$ and assumed here  to be diagonal and uniform only for simplicity of the presentation. 

In a NESS, the stationary distribution $p(\bf x)$ is time-independent, leading to a stationary current 
\beq
\jb(\xb)\equiv \beta D \Fb(\xb)p(\xb) - D \partial_{\xb} p(\xb)\equiv \nub(\xb)p(\xb)
\ee with the mean local velocity 
\beq
\nub (\xb)=\langle \dot \xb|\xb\rangle,
\label{eq:nu}
\ee
where $\langle ...\rangle$ denotes an average in the NESS. By going through similar steps as in Box 1 for the discrete case, the rate of physical entropy production becomes \cite{seif12,vdb15,peli21,shir23,seif25}
\beq\sigma =  \langle \nub^2(\xb)\rangle/D = \int d\xb \nub^2(\xb) p(\xb)/D.
\label{eq:sigma-con}\ee 

Even if this mesoscopic level, i.e., $i(t)$ or $\xb(t)$, is operationally accessible, determining $\sigma$ can still be challenging. In the discrete case,
one needs to infer the $p_i$s and the $k_{ij}$s from a long trajectory through measuring dwell times in all meso-states and the fluxes between those. In the continuous case, one has to measure the mean local velocity  either through a discretization of
(\ref{eq:nu}) or through
inference of the force and the stationary distribution\cite{fris20}.
In particular, if $\xb$ lives in two or more dimensions, there are subtle conceptual issues\cite{dieb22}. 
A reformulation of (\ref{eq:sigma-con}) in terms of derivatives of time-dependent correlation functions leads to the variance sum rule which, with some additional model assumptions, has been employed to estimate entropy production in the red blood cell membrane\cite{dite24}.

For both model classes, the expression for the entropy production, i.e., (\ref{eq:sigma}) or (\ref{eq:sigma-con}), can also be obtained by first identifying an entropy production along a trajectory $\gamma \equiv[i(t)]$ or $\gamma\equiv[\xb(t)]$ as
\beq
\Delta s\tot[\gamma]=\ln \{p[\gamma]/p[\tilde \gamma]\}.
\label{eq:sig-info}
\ee
through the log-ratio of the probabilities for a trajectory  and for its time-reversed counterpart $\tilde \gamma \equiv [i(T - t)]$ or $\tilde \gamma=[\xb(T-t)]$. Averaging then over all trajectories of length $T$ yields the mean total rate of entropy production
\beq
\sigma = \langle \Delta s\tot[\gamma] \rangle/T= (1/T)\sum_\gamma p[\gamma]\ln\{p[\gamma]/p[\tilde \gamma]\} .
\label{eq:sig-delta}
\ee
Crucially, as we will see later, the expression (\ref{eq:sig-info}) should not be taken as a universally valid identification of thermodynamic  entropy production. Instead, it reveals how this entropy production -- originally defined through entropy changes in the system and reservoirs -- can be recast in terms of trajectory probabilities for these two classes of systems.

When the mesoscopic level of description, as given by $i(t)$ and $\xb(t)$, is not fully accessible experimentally, the exact physical entropy production cannot be determined, in general. Consequently, the key challenge is to establish lower bounds on entropy production based on such partial information -- namely on data that involve some implicit or explicit coarse-graining over the inaccessible intermediate degrees of freedom.

Approaches meeting this challenge can broadly be classified into two groups. The first one follows the strategy to derive a bound on
$\sigma$ as given by (\ref{eq:sigma}) and (\ref{eq:sigma-con}).  This group comprises a variety of apparently unrelated methods based on the analysis of transitions between lumped states, on the thermodynamic uncertainty relation, and on the study of correlation functions.
A second strategy is to start with the expression (\ref{eq:sig-delta}) for $\sigma$ and to systematically coarse-grain the unobservable fine-grained trajectories $\gamma$ into observable coarse-grained ones $\Gamma$. 
 This approach leads to a hierarchy of bounds that get looser the more coarse-graining is applied.
In the following, we describe these two approaches, some of which are illustrated in Fig. 1, in more detail.

\section{Lower bounds in NESSs}

\subsection{State-lumping and optimal Markov models}
In the discrete case, state-lumping maps a set of mesoscopic states $\{i, j, k, l, \ldots\}$ onto a smaller set of coarse-grained states $\{\Ic, \Jc, \Kc, \ldots\}$, where each fine-grained state $i$ belongs to a lumped state $\Ic$, denoted as $i \in \Ic$, see Fig. 1. Given sufficient statistics from an (infinitely) long trajectory, an observer -- unaware of the underlying model or the lumping -- can still estimate the stationary distribution $p_{\Ic}$ and the flow (or rate) $\nu_{\Ic\Jc}$ between coarse-grained states. The connection between these coarse-grained quantities and the underlying mesoscopic ones is given by
\beq
p_\Ic=\sum_{i\in \Ic}p_i ~~~\textrm{and}~~~
\nu_{\Ic\Jc} = \sum_{i\in \Ic}\sum_{j\in \Jc}p_ik_{ij}.
\ee
We then get
\begin{eqnarray}
	\sigma&\geq& \sum_{\Ic\Jc}\sum_{i\in\Ic}\sum_{j\in\Jc} p_ik_{ij}\ln(p_ik_{ij}/p_jk_{ji})\nn\\
	&\geq&\sum_{\Ic\Jc} \nu_{\Ic\Jc}\ln(\nu_{\Ic\Jc}/\nu_{\Jc\Ic})\equiv \sigma\app \geq 0.
\label{eq:sigma-app}
\end{eqnarray}
The first inequality follows from (\ref{eq:sigma}) since transitions
within a coarse-grained state are not resolved by the observer.
The second one follows with the log-sum inequality
\beq
\sum_n a_n \ln(a_n/b_n)\geq \left(\sum_n a_n\right)\ln\left(\sum_n a_n/\sum_n b_n\right)
\ee valid for $a_n,b_n >0$.  The quantity $\sigma\app$, often called the 
apparent entropy production, thus yields an operationally accessible model-free lower bound on $\sigma$\cite{raha07,espo12,bo14}.

As Skinner and Dunkel\cite{skin21} have pointed out, this result can also be understood as the minimal entropy production of all discrete-state continuous-time Markov models that 
reproduce the stationary probability $p_\Ic$ and the fluxes $\nu_{\Ic\Jc}$ on the coarse-grained level. Following this approach, they have shown that if one additionally measures the rate $\nu_{\Ic\Jc\Kc}$ 
 at which two consecutive transitions between three
coarse-grained states happen, one can  ask for the optimal Markov model that has the smallest entropy production compatible with these observations. Even though this model cannot be obtained analytically anymore, it can be determined numerically. Applied to several sets of real-life data, they show that the latter approach can significantly improve the bound (\ref{eq:sigma-app}).

\subsection{Thermodynamic uncertainty relation (TUR)}
\subsubsection{Conceptual basis}
The TUR provides a universal lower bound on entropy production in a NESS through the mean of any current $\langle j\rangle $ and its variance $\var [j]$ recorded over a time $T$ as 
\beq
\sigma\tur \equiv 2 \langle j\rangle^2/\{T\var[j]\}
=\langle j\rangle^2/D_j
\leq \sigma,
\label{eq:tur}
\ee
where $D_j$ denotes the $T$--dependent diffusion coefficient  of this current, see Fig. 1.
 
For a system with discrete meso-states, the most general current reads
\beq
j[i(t)]=({1}/{T}) \sum_{i<j} \{n_{ij}[i(t)]-n_{ji}[i(t)]\}d_{ij} 
\label{eq:j}
\ee with weights $d_{ij}=-d_{ji}$ that are zero for unobserved transitions. The quantities $n_{ij}$ denote the number of transitions from $i$ to $j$ along a trajectory $i(t)$. Mean and variance of this current are
\beq
\langle j\rangle =\sum_{i<j}(p_ik_{ij}-p_jk_{ji})d_{ij}
\ee	
and
$\var[j]\equiv \langle \{j[i(t)]-\langle j\rangle\}^2\rangle =2D_j/T,
$
respectively.

For an overdamped Langevin dynamics, a general current
can involve a weight function $g(\xb)$  and then reads
\beq
\jb[\xb(t)]=(1/T)\int_0^Tdt g(\xb^t) \dot \xb(t)
\ee with mean $\langle \jb\rangle = \langle g(\xb) \nub(\xb)\rangle
=\int d\xb g(\xb)\nub(\xb)p(\xb)$.

Based on extensive numerics and analytical limit cases, the TUR (\ref{eq:tur}) has first been conjectured in the long-time limit ($T\to \infty$)\cite{bara15}. It has been proven  using techniques from large deviation theory\cite{ging16}. The finite-time version, which is relevant for applications, has been conjectured in \cite{piet17} and proven in \cite{horo17}. Alternative proofs have been based on generating functions \cite{dech17}, on exploiting the Hilbert space structure of observables\cite{fala20},  on stochastic calculus \cite{dieb23}, 
and on a more general fluctuation-response relation \cite{asly25}.

The TUR  can be saturated close to equilibrium, and, even far from equilibrium, in the short-time limit $T\to 0$ as suggested in \cite{mani20} and proven in \cite{otsu20,kim20}.
Optimization with respect to the current \cite{busi19} and, if several currents are available, tighter versions \cite{dech21} using their covariances have been discussed.

Crucially, for using the TUR as a lower bound on entropy production, information about the underlying system is not required beyond the assumption that, on a deep underlying
level, the system follows a Markovian dynamics on
discrete states or an overdamped Langevin dynamics. For
an  underdamped Langevin dynamics in more than one-dimension it can be broken \cite{piet21}. In the one-dimensional case, numerical work \cite{fisc19} indicates its validity in the long-time limit. However, a proof for this conjecture has not yet been found.

\subsubsection{Biophysical applications}

As a striking consequence of the TUR, the thermodynamic efficiency $\eta$ of a molecular motor moving with mean velocity $v$ against an externally applied force $f$  can be bounded by \cite{piet16b} 
\beq
\eta\equiv fv/[fv + \sigma/\beta]\leq 1/[1+v/(fD_v\beta)] .
\ee
Here, the definition recalls that efficiency is the ratio between power output and input, while their difference is given by the thermodynamic cost, i.e., by the entropy production. The latter is bounded by the TUR applied to the fluctuating velocity as a current with dispersion $D_v$. This bound contains only experimentally accessible quantities. Applied to experimental data for kinesin moving along a microtubule \cite{viss99}, this expression  yields upper bounds on the efficiency with which this molecule transforms chemical energy in the form of ATP to mechanical power working against the externally imposed load \cite{seif17}. Again, the sole assumption behind this application is the idea that at some deep level the complex consisting of the molecular motor and the attached bead through which the force is applied can be described by a Markovian dynamics on discrete or overdamped continuous states. Applications of the TUR to theoretical models of such motors show that this universal bound may become quite loose in specific cases\cite{hwan18}. The precision-cost trade-offs of further biomolecular processes in the light of the TUR have been explored in \cite{song21}.

For active cell membranes, Manikandan et al\cite{mani24} used the short-time limit of the TUR to get a spatially and temporally resolved lower bound on entropy production from experimental data of flickering. Even though this value is tiny --  about $10^{-2}\kbs$ for a membrane patch of $4 \mum^2$ --, this study demonstrates that this model-free approach can be applied to real-life data. 

Some applications of the TUR to living systems require additional model assumptions. In\cite{rold21},
an experimentally supported model for the spontaneous hair-bundle oscillations in mechanosensory hair cells from the ear of the bull-frog was simulated to predict
that an entropy production rate of about $10^3\kbs$ could be measured as lower bound from the TUR if one had access to two variables measuring the tip position and the transduction current. The swimming of sperms due to the beating of their tails has been analyzed using the TUR\cite{magg23}.
The resulting value $\sigma\tur \simeq 100\kbs$ is still about five orders of magnitude lower than what is expected due to viscous dissipation. A model for strongly coupled motors has been argued to resolve this apparent discrepancy.

\subsection{A bound from observed ticks}
The TUR requires the observation of a current, i.e.,  an observable that is odd under time reversal. Suppose a non-equilibrium system produces just a sequence of $n(T)$ ``ticks''  in a time $T$ like the spike train of a neuron. The precision of such a time series consisting of even observables can be characterized 
by its Fano factor 
\beq
F\equiv \lim_{T\to \infty} \var[n(T)]/\langle n(T)\rangle
\ee 
Since even an equilibrium system can produce ticks, there is a threshold on $F$ above which no driving is required. 
For ticks that result from transitions in an underlying discrete-state Markov network, Pietzonka and Coghi \cite{piet23} have derived a lower bound on entropy production
$\sigma \geq \bar\sigma(F)
/\langle \tau\rangle. 
$
Here, $\langle \tau\rangle$ is the mean time between ticks and the function
$\bar\sigma(F)$ 
is given by an implicit analytical expression that comes in two variants. 
If in the underlying network, a transition through a link generates a tick irrespective of its direction, then the threshold for  $F$ is at 1, i.e. $\bar \sigma(F)=0 $ for $F\geq 1$. If some transitions lead to a tick when traversed in only one direction, the bound  becomes smaller and the threshold is at $F=1/2$. As in the case of the TUR, for $F\to 0$, the entropy production diverges. Thus, infinite precision of ticks implies an infinite thermodynamic cost in the system generating the ticks. The application to an experimentally measured spike train leads to an inferred cost of up to about 30 $\kbT$ per spike \cite{piet23}.

\subsection{Bounds from correlation functions and oscillations}

In equilibrium, correlation functions $\langle A(t) B(0)\rangle$
are symmetric. Any
$\langle 
A(t)B(0)\rangle \not = \langle B(t) A(0)\rangle$
is a signature of non-equilibrium. How to turn this asymmetry into a quantitative model-free lower bound on $\sigma$, however, is non-trivial.

For an overdamped Langevin dynamics, the
 auto-correlation function $\langle A(t)A(0)\rangle$ for a bounded observable $A(\xb)$ with
$A_\textrm{max}\geq A \geq A_\textrm{min}$ and $\langle A\rangle=0$ can yield a lower bound on $\sigma$. First, a mean correlation or relaxation time is
extracted through
\beq\tau_A\equiv \int_0^\infty dt \langle A(t)A(0)\rangle/\langle A^2\rangle .
\ee
 The short time fluctuations become $\langle [A(t)-A(0)]^2\rangle = 2 D_A
t + O(t^2)$.
 With these two quantities, one has\cite{dech23}
\beq\sigma\geq 
[4/(A_\textrm{max}-A_\textrm{min})^2][\langle A^2\rangle/\tau_A-D_A].\ee
Especially close to equilibrium, this bound may become weak and, depending on the observable, even trivial since the right-hand side may become negative. 

Oscillations in  an auto-correlation function $\langle A(t)A(0)\rangle$
that involves states or overdamped position variables necessarily require non-equilibrium conditions since in equilibrium such an auto-correlation function decays monotonically. The more coherent or persistent such correlations are, the larger is the entropy production per oscillation. If this correlation function shows a number $\N$ of coherent oscillations, i.e., if after $\N$ oscillations the amplitude of the correlation function has decreased by a factor $e\simeq 2.718$,  the mean entropy production per oscillation $\Delta S\osc$  is
conjectured to be bounded by\cite{ober22}
\beq \Delta S\osc \geq 4\pi^2\N
\ee 
independent of the details of the underlying Markovian dynamics on discrete or overdamped continuous states. In terms of the complex dominant eigenvalue $\l$ of the generator of the dynamics, this yet unproven conjecture formally becomes $\sigma\geq \Im^2 (\l)/\Re (\l)$ for $\N>1/(2\pi)$. In the weak-noise limit of continuous systems, the bound can be proved\cite{reml22,sant25,naga25a}. Related bounds involving oscillations have been discussed in 
\cite{cao15,bara17,mars19,ohga23,kolc23}.

\section{Coarse-graining trajectories}

\subsection{A master relation for a universal lower bound}
Quite generally, coarse-graining corresponds to a many-to-one mapping of fine-grained trajectories $\gamma$ on meso-states (with $\gamma=[i(t)]$ and
$\gamma=[\xb(t)]$ in the discrete and in the continuous case, respectively)
to  coarse-grained ones, denoted by $\Gamma$. We write this mapping as 
$
\Gamma: \gamma\mapsto \Gamma[\gamma] .
$ The path weight, i.e., the probability density of a coarse-grained trajectory $\Gamma$ follows
from the one of all fine-grained $\gamma$s contributing to $\Gamma$ according to
\beq
p[\Gamma]=\sum_{\gamma\in \Gamma}p[\gamma] .
\label{eq:p-gaGa}
\ee

If the coarse-graining satisfies an additional key condition related to time-reversal symmetry, a universal lower bound on physical entropy production can be derived that depends solely on the coarse-grained trajectories. Specifically, time reversal must induce a one-to-one mapping between coarse-grained trajectories, $\Gamma\mapsto \tilde \Gamma[\Gamma]$, that commutes with coarse-graining, i.e., 
\beq
\Gamma[\tilde \gamma]=\tilde \Gamma [\Gamma[\gamma]]
\label{eq:cg-timerev}
\ee has to hold.

In a NESS, we can then rewrite (\ref{eq:sig-delta}) as
\begin{eqnarray}
\sigma &=& \frac{1}{T}\sum_\Gamma \sum_{\gamma \in \Gamma}p[\Gamma]p[\gamma|\Gamma]
\ln\left[p[\Gamma]p[\gamma|\Gamma]/p[\tilde\Gamma]p[\tilde \gamma|\tilde\Gamma]\right]\nn\\
&=&\frac{1}{T}\left[\sum_\Gamma p[\Gamma]\ln\frac{p[\Gamma]}{p[\tilde \Gamma]} + \sum_\Gamma p[\Gamma] \DKL[p[\gamma|\Gamma]||p[\tilde \gamma|\tilde \Gamma]]\right] 
\end{eqnarray}
with the Kullback-Leibler divergence $\DKL[p(x)||q(x)] \equiv \sum_xp(x) \ln[p(x)/q(x)] \geq 0$ for any two distributions $p(x)$ and $q(x)$. One thus gets the lower bound
\beq\sigma\geq \sigma\cgr\equiv \frac{1}{T}
\sum_\Gamma p[\Gamma]\ln\frac{p[\Gamma]}{p[\tilde \Gamma]}
\label{eq:sigma-cg} 
\ee
that has been introduced in \cite{kawa07,gome08a,gome08b} to stochastic thermodynamics.
 
The inequality (\ref{eq:sigma-cg}) crucially requires that time reversal and coarse-graining commute as emphasized in \cite{hart21a}. Otherwise, it may happen that $\sigma\cgr>\sigma$ and that $\sigma\cgr$ is non-zero even in equilibrium. For an instructive discussion of this issue, see \cite{mart19,hart24,bisk24}. In the following, we will focus on 
coarse-graining schemes that are thermodynamically consistent since they comply with the condition
(\ref{eq:cg-timerev}).

\subsection{Thermodynamically consistent coarse-graining}
\subsubsection{State-lumping trajectories}
State-lumping, as introduced above for discrete states,
yields a  conceptually simple yet widely used, practically important and thermodynamically consistent  form of coarse-graining. It can also be applied to continuous states which are then mapped onto a smaller set of coarse-grained states  $\{\bf X\}$ or $\{\Ic\}$  according to
$ {\bf x} \mapsto \textbf{X} ~~~\textrm{or}~~~{\bf x \mapsto  \Ic}$. The
first version applies if, e.g., only a subset of coordinates are observed. The second version applies in principle to any experimental recording of a continuous degree of freedom due to the inevitable finite resolution of imaging data caused by the pixel size.
Since states and positional degrees of freedom are even under time reversal, this type of coarse-graining indeed commutes with time reversal and, hence, is thermodynamically consistent.  
 
\subsubsection{Transition-based coarse-graining}

Here, we assume that only a subset of the transitions from the Markovian
discrete-state fine-grained dynamics is observable. We denote this set by $\{I,J,K,...\}$ where each capital letter denotes a fine-grained transition, e.g., $ I =i\to j$.
A coarse-grained trajectory then consists of a sequence of transitions and waiting times between these transitions, $\Gamma=(I_0,\tau_1,I_1,\tau_2,...)$.

Crucially, transitions are ``odd'' under time reversal since a transition
$I\equiv i\to j$ 
becomes $\tilde {I} \equiv j\to i$. 
Still, if the mapping indicating time reversal on the coarse-grained level is so defined, coarse-graining and time reversal commute in the sense of (\ref{eq:cg-timerev}). Therefore, the bound (\ref{eq:sigma-cg}) holds true for coarse-grained trajectories that contain only a subset of fine-grained transitions provided that for each observable transition its time-reversed partner is also observable.

\subsubsection{Temporal coarse-graining}
In principle, any experimental recording of a continuous-time trajectory  leads to a discrete-time trajectory due to a finite-temporal resolution. 
Likewise, one could imagine to record a continuous-time system stroboscopically in regular intervals\cite{cisn23,baue25}. In these cases, the coarse-grained trajectories are of the type
$
\Gamma=
(a_0,a_1,a_2,...)
$
where the $a_i$ are either discrete or continuous fine-grained -- or already coarse-grained -- states. Again, if the
time-reversed trajectory has been identified thermodynamically consistently on each coarse-graining step, the bound (\ref{eq:sigma-cg}) holds true
when evaluated with such  coarse-grained trajectories.

A variant of temporal coarse-graining is the mapping 
\beq
\Gamma=(a_0,\tau_0,a_1, \tau_1,a_2, \tau_2,...) \mapsto 
 \Gamma'= (a_0,a_1,a_2,...)
\label{eq:gammap}
\ee where the intermediate dwell or waiting times between the events $a_i$ are ignored or deleted.

\subsection{A fundamental challenge and one possible solution}
A key challenge for the evaluation of the lower bound (\ref{eq:sigma-cg}) is the required averaging over all coarse-grained trajectories, which remains difficult because the probability of observing time-reversed trajectories decreases exponentially with trajectory length $T$.

For a discrete coarse-grained trajectory $\Gamma=(a_1,a_2,...)$, one strategy introduced in \cite{rold10,rold12} is to break the time-series into $k$-tupels, which yields a $k$-dependent estimate of the entropy production that, in the limit $k\to\infty$, approaches (\ref{eq:sigma-cg}).
Specifically, one has to form the Kullback-Leibler divergences
\beq \DKL_k \equiv \sum_{a_1,...,a_k} p(a_1,...,a_k) \ln\frac{p(a_1,...,a_k)}{p(a_k,...,a_1)},
\label{eq:DKL}
\ee where $p(a_1,...,a_k)$ is the probability distribution for the $k$-tupel $(a_1,...,a_k)$ in a long  trajectory.
The mean rate of Kullback-Leibler divergence per datum of this series is then given by
\beq\sigma\dkl\equiv \lim_{k\to \infty}(1/k)\DKL_k=\lim_{k\to \infty} [\DKL_{k+1}-\DKL_{k}].
\ee
 If the events arise from a Markov chain, $\DKL_k=(k-1)\DKL_2$ for $k\geq3$. Likewise, for an $n$-th order Markov chain, the second limit is  reached for $k=n+1$.
 If the coarse-graining leading to (\ref{eq:sigma-cg}) has been consistent,
$\sigma\dkl/\langle \tau\rangle$ provides a lower bound on the rate of physical entropy production, where $\langle \tau\rangle$ is the mean time interval between data points in the series.

\section{Bounds from Waiting-time distributions}

Waiting-time distributions between a subset of transitions in the underlying Markov network can yield a bound on its entropy production.
This idea has shown up in \cite{mart19,skin21a,ehri21} and has been developed into a
comprehensive framework independently in \cite{vdm22} and \cite{haru22}. 

Specifically, we assume that 
an observer has access to a certain set of
transitions $\{I,\tilde I,J,\tilde J, K, \tilde K,... \}$, where for each observable transition the time-reversed counterpart is observable as well, see Fig. 1. In a NESS,
the observer can thus generate waiting-time distributions
\beq\psi_{IJ}(t)=p(J,t|I) ,
\label{eq:wtd}
\ee which are the probability density that, given transition $I$ happens at time 0, the next observed transition is $J$ at time $t$ where $J=I$ and $J=\tilde I$ is possible in general. The normalization is given by
\beq\sum_J\int_0^\infty dt \psi_{IJ}(t)\equiv\sum_J p_{IJ} = 1,
\label{eq:norm}
\ee
where $p_{IJ}$ is the probability that transition $J$ follows $I$.

A coarse-grained trajectory $\Gamma$ -- starting and ending with observed  transitions $I_0$ and $I_n$, respectively -- thus becomes
\beq\Gamma = (I_0,\tau_1,I_1,\tau_2,I_3,\tau_3, ...,I_n) 
\label{eq:G1}
\ee
with the time-reversed counterpart
\beq\tilde \Gamma = (\widetilde {I_n}, \tau_n,\widetilde {I_{n-1}}, \tau_{n-1},\widetilde {I_{n-2}},\tau_{n-2},...,\widetilde {I_0)}.
\label{eq:G2}
\ee
Since on the level of the meso-states $\{i\}$ the dynamics is Markovian, the weight of the coarse-grained trajectory can be factorized into weights of the
paths between these transitions. They are called Markovian events since in these instances the future dynamics becomes independent of the past, i.e., independent of what happened before this observed transition. 
The paths between these Markovian events are dubbed Markovian snippets\cite{vdm22b}. 

In a NESS, the weights for the two trajectories (\ref{eq:G1}) and (\ref{eq:G2}) can thus be expressed by the waiting-time distributions as
\beq
p[\Gamma]= p(I_0)\prod_{j=1}^n p(I_j,\tau_j|I_{j-1})
=p(I_0)\prod_{j=1}^n \psi_{I_{j-1}I_j}(\tau_j)
\label{eq:G11}
	\ee
and
\beq p[\tilde \Gamma]= 
p(\widetilde{I_n})\prod_{j=1}^n p(\widetilde{I_{j-1}},\tau_j|\widetilde{I_{j}}) 
=p(\widetilde{I_n})\prod_{j=1}^n \psi_{\widetilde{I_{j}}\widetilde{I_{j-1}}}(\tau_j),
\label{eq:G22}
	\ee
respectively.
Inserting this into  (\ref{eq:sigma-cg}), taking into account that in the limit $T\to \infty$ boundary terms vanish, and noting that in a NESS the
transition $I$ (with $I=i\to j$) happens with a rate
$\nu_I=p_ik_{ij}$ leads to the lower bound\cite{vdm22}  
\beq\sigma\wtd \equiv \sum_{IJ}\int_0^\infty dt\nu_I\psi_{IJ}(t)\ln\frac{\psi_{IJ}(t)}{\psi_{\tilde J \tilde I}(t)} \leq \sigma.
\label{eq:sig-wtd}
\ee
With 
the log-sum inequality applied to the time integration and using (\ref{eq:norm}) one gets the weaker bound 
\beq
0\leq \sigma\emc\equiv \sum_{IJ}\nu_I p_{IJ}\ln\frac{p_{IJ}}{p_{\tilde J\tilde I}}\leq \sigma\wtd,
\label{eq:sig-emc}\ee
which is the entropy production of the embedded Markov chain obtained from the sequence of transitions by ignoring the waiting times. 
For an underlying unicycle, the bound $\sigma\emc$ reproduces the exact mean entropy production rate $\sigma$ even if only one link is observed. For an underlying multi-cyclic system, $\sigma\emc$ becomes exact if at least one link from each fundamental cycle is observed. In both cases, the ratio of waiting-time distributions is time-independent such that $\sigma\wtd=\sigma\emc=\sigma\cite{vdm22}.$ The conjecture that in general $\sigma\wtd\geq \sigma\tur$ has not yet been proved.

Further
generalizations of this approach based on Markovian events and snippets are discussed and illustrated in Box 2.  

\section{Entropy production along individual coarse-grained trajectories}
The estimators discussed so far yield lower bounds on the mean entropy production in a NESS. Since stochastic thermodynamics provides a unique expression for entropy production along individual fine-grained (mesoscopic) trajectories \cite{seif05a}, which is given by (\ref{eq:sig-info}) for a dynamics on discrete and overdamped continuous states, the question arises whether or not an entropy production along individual observed coarse-grained trajectories can be identified consistently. Indeed, the relations (\ref{eq:sig-delta}) and (\ref{eq:sigma-cg}) seem to suggest the identification of a trajectory-dependent entropy production along a coarse-grained trajectory according to
$\Delta S\tot[\Gamma]\equiv\ln\{p[\Gamma]/ p [\tilde \Gamma]\}$. However, the presence of the boundary terms in (\ref{eq:G11}) and  (\ref{eq:G22}) as well as
a similar term appearing for underdamped dynamics, see (\ref{eq:box3-7}) in Box 3,  indicate that such an assignment cannot universally be correct. In fact, a thermodynamically consistent identification requires one modification and one more restriction \cite{degu23}. First, the correct identification is
\begin{eqnarray}
	\Delta S\tot[\Gamma]&\equiv& \ln \frac{p[\Gamma|\Gamma^0]p(\Gamma^0)}{p[\tilde \Gamma|\tilde\Gamma^0]p(\Gamma^T)}\nn\\
	&=&\ln \{p[\Gamma]/p[\tilde \Gamma]\}+\ln [p(\widetilde{\Gamma^T})/p(\Gamma^T)] .
	\label{eq:sigma-me}
\end{eqnarray}
Second, this identification is guaranteed to be consistent only if additionally the coarse-grained trajectory starts and ends with  Markovian events $\Gamma^0$ and $\Gamma^T$, respectively. 
If these Markovian events are even under time reversal -- which they are if the event comprises being in  a state  or being at a certain position in space -- then the boundary term in (\ref{eq:sigma-me}) vanishes and one recovers what one would have naively expected. However, if the Markovian event is odd, since, e.g., it
consists of the observation of a transition between two meso-states (or of the velocity for underdamped dynamics)  then the boundary term in (\ref{eq:sigma-me}) becomes crucial for a consistent assignment of  the coarse-grained entropy production. 

The thermodynamic consistency of the identification (\ref{eq:sigma-me}) is guaranteed since the relation
\beq
\langle \Delta s\tot[\gamma]|\Gamma\rangle \geq \Delta S\tot[\Gamma]
\label{eq:IFT-cg0}
\ee
 shows that the coarse-grained entropy production is not bigger than the original entropy production averaged over  all fine-grained mesoscopic trajectories contributing to the coarse-grained one. This inequality follows from the integral fluctuation relation \cite{degu23,ferr25}
\beq
\langle \exp\{-\Delta s\tot[\gamma]\}|\Gamma\rangle =
\exp\{-\Delta S\tot[\Gamma]\}
\label{eq:IFT-cg1}
\ee that can easily be proved based on the identification (\ref{eq:sigma-me}). 
Upon averaging with $p[\Gamma]$, one obtains from (\ref{eq:IFT-cg0}) and (\ref{eq:IFT-cg1}) the second law on the coarse-grained level 
\beq
\langle \Delta s\tot[\gamma]\rangle \geq \langle \Delta S\tot[\Gamma]\rangle \geq 0. 
\ee

\section{ Relaxation and time-dependent driving}

So far, we have discussed non-equilibrium steady states, which are characterized by a time-independent stationary distribution and  a constant mean entropy production rate. Systems that are initially out of stationarity -- or, under equilibrium conditions, not yet equilibrated -- relax into a NESS or an equilibrium state, respectively. During this relaxation, the mean instantaneous entropy production rate $\sigma(t)$ becomes time-dependent. Likewise, systems driven by time-dependent protocols $\l(t)$ show an instantaneous $\sigma(t)$.

The general expressions for the entropy production rate, given by (\ref{eq:sigma}) and (\ref{eq:sigma-con}) for discrete and overdamped continuous systems, respectively, remain valid for the instantaneous one if the distributions, transition rates, and forces are replaced by their time-dependent counterparts\cite{seif25}. This holds under the assumption that the dynamics of the system remains Markovian at the mesoscopic level -- that is, the microscopic degrees of freedom relax faster than the external driving changes.

Several of the methods described in the sections above have been generalized to yield lower bounds on the instantaneous rate of entropy production or on its time-integrated version corresponding to the mean entropy change of a process as we will now discuss. 
\subsection{Thermodynamic uncertainty relation}

The TUR can be generalized
to periodically driven systems. For discrete systems,  the rates are then periodic functions with a frequency $\Omega$, whereas for overdamped dynamics the force becomes periodic in time. In the corresponding periodic steady state, the entropy production, the mean current and the long-term diffusion coefficient  become functions of $\Omega$ denoted by
$\sigma(\Omega),J(\Omega)$ and $D_J(\Omega)$, respectively.
The TUR then reads\cite{koyu19a}
\beq\sigma(\Omega)\geq [J(\Omega)-\Omega J'(\Omega)]^2
/D_J(\Omega),
\ee
with $J'(\Omega)\equiv \partial_\Omega J(\Omega)$.
In contrast to the TUR for NESSs, even time-symmetric observables such as
\beq A[i(t)]\equiv (1/T)\int_0^T dt \sum_i\delta_{i,i(t)}A_i(t)
\ee yield a bound on entropy production. Here, $A_i(t)$ can be any periodic state function  like, e.g.,  the free energy of meso-state $i$. In this case, one gets
\beq\sigma(\Omega)\geq [\Omega A'(\Omega)]^2/D_A(\Omega) ,
\ee with  the long-time mean $A(\Omega)$ 
and diffusion coefficient $D_A(\Omega)$ of $A$, respectively.
It is thus sufficient to measure how the mean current
-- or the mean state observable -- depends on the frequency of driving. Knowledge of the transition rates (or of the force) is not required for evaluating this model-free lower bound.

 For systems that are relaxing from an arbitrary initial distribution and for systems that are driven by  an arbitrary time-dependent protocol, operationally accessible generalizations of the TUR have been derived\cite{liu19,koyu20,dieb24}. An instructive presentation of the hierarchy behind these bounds is given  in \cite{kwon24}. 

\subsection{Bounds from waiting-time distributions} The lower bounds based on waiting-time distributions can straightforwardly be generalized to
time-dependent driving. As a main ramification, the waiting-time distributions acquire a second time argument as in $\psi_{IJ}(t)\to \psi_{IJ}(t,t_I)$ where $t_I$ is the time when event $I$ occurs and $t=t_J-t_I$. Likewise, the identification of a coarse-grained entropy production along trajectories is possible even in the time-dependent case. For the general theory, and an application to paradigmatic experimental data on the unfolding of a small peptide, see\cite{degu24}. The special case of periodic driving has been investigated in \cite{maie24,haru24a}.

\section{Bridging theory and experiments}

Conceptually -- and also when discussing experiments --
one should distinguish qualitative signatures that indicate non-equilibrium from its quantification for which entropy production arguably is the primary measure.

On a qualitative level, any violation of detailed balance through the observation of a non-vanishing current or an asymmetry in a cross-correlation indicates non-equilibrium\cite{batt16,baca23}. Turning this into a quantitative measure -- or rather a lower bound on entropy production -- typically requires to focus on a few modes or principal components and then to analyze the fluxes between them in an expression like (\ref{eq:sig-emc}) as it has been studied for the cytoskeleton\cite{sear18} and even for human brain activity\cite{lynn21}. The
typically rather small values -- of the order of 0.1$\kbs$ in the latter experiment -- arise from the rather low dimensional effective phase space that is experimentally accessible. 

Likewise, any  violation of the fluctuation-dissipation theorem -- as found for
hair-bundle cells \cite{mart01}, for cytoskeleton networks\cite{mizu07}, for the flickering of red-blood-cells\cite{turl16}, and  for the brain\cite{deco23} -- indicates a non-equilibrium system. Relating such a violation quantitatively to entropy production is non-trivial. For systems following a Langevin dynamics, it has been achieved  through the Harada-Sasa relation \cite{hara05} that has been applied to experimental data of molecular motors \cite{toya10,arig18}. 
As an aside, note that strictly speaking any approach based on a fluctuation-response-type relation is invasive, since it requires to measure the response to an external perturbation.

Only a few references so far have demonstrated the application of the bounds based on the TUR and on waiting-time distributions to experimental data as cited above. Indeed, attempts to infer  or even bound entropy production from measurements remain scarce since obtaining trajectories with sufficient statistics under non-equilibrium conditions is inherently challenging. Previously, such data have been successfully generated for colloidal particles in harmonic, periodic, and bistable potentials generated by optical tweezers, as reviewed in \cite{seif12,cili17,mart17}, and -- with even better statistics -- for electronic devices that can still be described by a classical master equation \cite{peko19}. In both types of systems, however, the relevant degrees of freedom are directly accessible, making the use of these more recent conceptual tools largely unnecessary.
 
 Bio-systems present greater challenges in identifying sufficient statistics. Representative trajectory data from single-molecule experiments include studies of linear and rotary molecular motors \cite{toya10,arig18}, as well as unfolding of proteins and nucleic acids\cite{stig11,alem15}. A further promising field is  fluorescence resonance energy transfer (FRET); for reviews, see \cite{roy08,lern18,maza19,nett24}. Theoretical efforts to infer thermodynamic quantities from FRET-data have only recently begun \cite{gode23,voll24,voll24a,song24}.


A complementary approach to infer entropy production could involve  
calorimetric measurements of heat. The current limit in an aqueous environment is the picowatt range, see, e.g.,  Ref.\cite{fost23} for an experiment on cytoskeletal material. The ability to resolve the dissipation arising from single molecular reaction events -- expected to be several $\kbT$  -- is still orders of magnitude off since 1pW$\simeq 2.4 \times 10^8\kbT
$/s.  
Moreover, in general, dissipated heat differs from genuine entropy production in at least two cases. First, whenever chemical reactions are involved, entropy production additionally includes an entropy change in the surrounding solution. Second,  
in relaxing or time-dependently driven systems, entropy production contains a contribution from a changing system entropy. The bounds discussed here refer to all contributions to entropy production -- not just to the exchanged heat.
\section{Further challenges and perspectives}

\subsection{A caveat regarding trajectories obtained from uncontrolled or thermodynamically inconsistent coarse-graining}
Throughout this review, we have emphasized the essential role of enforcing thermodynamic consistency at each stage of coarse-graining in order   to derive meaningful lower bounds on entropy production. Given an ensemble of observed, apparently stationary trajectories $\{\Gamma\}$, it is not obvious whether they arose from an effective, thermodynamically consistent coarse-graining of an underlying Markov model. Formally, given sufficient statistics, one can still determine
\beq\mathcal{S} \equiv \sum_\Gamma p(\Gamma) \ln[p(\Gamma)/p(\tilde \Gamma)] \geq 0 
\label{eq:sig-info2}
\ee
for any one-to-one mapping $\Gamma \mapsto \tilde \Gamma[\Gamma]$ mimicking time reversal. This non-negative quantity is positive whenever original and ``time-reversed'' trajectories have a different statistics. In general, however, the thus defined proxy for entropy production is not a lower bound on the physical entropy production. A sufficient condition for it to be one is that the time reversal on the coarse-grained level commutes with the time reversal of the underlying fine-grained trajectories. 

An instructive example for coarse-graining that does not obey this condition is milestoning\cite{elbe20}. Here, a coarse-grained trajectory is characterized by the last milestone it has passed and the time elapsed since then.
A comprehensive theory of how to extract bounds on entropy production from such data has not yet emerged even though some practical strategies have been explored in case studies\cite{blom24}. If, e.g., the milestones consist of lumped fine-grained states, a consistent bound can be obtained by focusing on the sequence of milestones and ignoring the waiting times between them as in (\ref{eq:gammap}).

\subsection{Thermodynamic versus informatic entropy production: Active particles}

A related issue concerning time reversal arises in the context of active particles. The self-propulsion of these particles in an aqueous environment is fuelled by the consumption of free energy generated by local chemical reactions. Rather than modeling those, one might want to infer entropy production directly from a postulated phenomenological stochastic dynamics on a coarse scale by then comparing the probabilities of forward and time-reversed trajectories as in (\ref{eq:sig-info2}).

The limitations and subtleties of this approach can be exemplified by the paradigmatic case of an active Brownian particle.
 Within a phenomenological model and for simplicity in its overdamped version,  such a  particle with mobility $\beta D$ in a potential $V(\xb)$ follows the dynamics\cite{roma12,chau14}
\beq\dot\xb(t)=u\nb(t) - \beta D\nablab V(\xb^t) +  \zb(t)	 
\label{eq:lactive}
 .\ee
 The self-propulsion with strength $u$ acts along the director $\nb$ that undergoes Brownian rotational motion. If this self-propulsion is interpreted as a force, then the director remains unchanged under time reversal. If instead it is viewed as a self-propelling velocity, it should be reversed. These two interpretations inserted into (\ref{eq:sig-info}) lead to different expressions for entropy production \cite{fodo16,mand17,piet17b}. At the level of a phenomenological Langevin equation, it remains undetermined which formulation correctly represents the physical entropy production.
 
A definite answer requires to model the physical mechanism behind the self-propulsion more explicitly. For two major classes of active particles, active Brownian particles and active Ornstein-Uhlenbeck particles, this has been achieved in \cite{piet17b} and \cite{frit23}, respectively, through a thermodynamically consistent lattice model in which the propulsion is caused by a simple chemical reaction.  One further key result arising from the continuum limit of such models is the observation that the description by a Langevin dynamics (\ref{eq:lactive}) contains an implicit coarse-graining since it does not resolve the separate contributions of thermal fluctuations and individual reaction events. Consequently, a thermodynamically consistent Langevin equation should include anisotropic friction and noise arising from the chemical reactions \cite{piet17b,gasp17}. An off-lattice version of a thermodynamically consistent model is discussed in\cite{bebo25} which contains a comprehensive bibliography of related work.

The insights gained for this paradigmatic system have lead to the conclusion that the entropy production resulting from a naive application of (\ref{eq:sig-info2}) to trajectories with whatever
definition of time-reversed trajectories should not necessarily be taken as an estimate of thermodynamic entropy production -- or a lower bound to it -- 
but rather as one quantitative measure of irreversibility better called informatic entropy production \cite{fodo22}.
This quantity is guaranteed to be a lower bound on physical entropy production if it arises, at least in principle, from coarse-graining an underlying model that  respects the condition (\ref{eq:cg-timerev}).

\subsection{Bulk systems: Local and scale-dependent entropy production}

The systems discussed so far refer essentially to single particles and to single  molecules or a few of them. In principle, these approaches are applicable to many-particle systems provided one has generated sufficiently detailed trajectory data. The case study of a driven lattice gas\cite{koyu22} has shown that the TUR-bound remains non-trivial in the thermodynamic limit.

A complementary description of a many-particle bulk system focuses on densities and currents and, more generally, on fields that follow a stochastic equation of motion\cite{dean96}. From the weight for the space-time history of these fields one may extract a time- and space-local expression for informatic entropy production that is a lower bound on the physical one with the caveats concerning correct time reversal just mentioned. 
Such an approach has been  pursued quite intensively recently in the active matter community, mostly in theory as reviewed in \cite{obyr22,nard17}, and also in an experiment\cite{ro22}. 

For general networks, an important issue is how entropy production scales with the scale of coarse-graining. The scaling observed in several studies \cite{yu21,cocc22,yu24} should not necessarily be taken as a signature of physical entropy occurring on such scales according to \cite{schw25}. 
A better understanding whether and how entropy production scales especially in inhomogeneous systems will be important to get theoretical estimates, based on necessarily coarse-grained experimental data, closer to the real values.
\subsection{A few extensions}


Throughout, we have emphasized that the bounds discussed here are universal, i.e.,  model-free. Obviously, if more information about an underlying network is available, such as its topology, then coarse-grained data can be used in more specific expressions that lead to tighter bounds or even the correct value of the entropy production, see, e.g., \cite{teza20,nitz23,igos25}.

Apart from entropy production, there are further, more refined thermodynamic quantities. For discrete networks, the cycle affinity $\mathcal{A}_c=\kbT\ln\prod_{i\in C}(k_i^+/k_i^-)$, determined by the ratio of all forward and backward rates in a cycle $C$, is a measure of the external driving associated with this cycle. This quantity normalized with the number of states of the cycle enter a sharper version of the TUR\cite{piet16} and a bound on the asymmetry of cross-correlations\cite{lian23,ohga23}.


An implicit assumption so far has been that if a state and a transition is observable, this observation and the subsequent coarse-graining will be error-free. Investigations of faulty coarse-graining and of the role of random
transition-detection blackouts have just started\cite{vdm25,bao25,maie25b}. Another restriction of experimental work may be the fact that reverse transitions can be very rare or even absent in a finite trajectory. A theoretical scheme to mitigate this effect has been suggested 
in\cite{baie24}. The role of finite statistics for the bounds based on the TUR and waiting-time distributions has been explored in\cite{frit25}.

Finally, hydrodynamics can affect entropy production both through externally imposed flow\cite{spec08} and through hydrodynamic interactions\cite{gasp20,das25,gole25}.

\subsection{The two faces of entropy bounds}

The approach discussed in this review is based on inequalities that relate entropy production on the left-hand side with a function of some coarse-grained data on the right-hand side. Since entropy production is the thermodynamic cost of a process, a complementary perspective on such an inequality is that it gives the minimal cost of achieving whatever the right-hand side expresses. If this right-hand side characterizes a desired quantitative outcome (or feature
or benchmark) of a process,
then the inequality gives the minimal thermodynamic cost of achieving this outcome. 
 Within such an interpretation, e.g.,  the TUR gives a lower bound on the thermodynamic cost of reaching a certain temporal precision. For heat engines,  a variant of the TUR quantifies the trade-off between power, efficiency and constancy \cite{piet17a}.
Speed limits quantify the cost of transforming a given initial distribution into a prescribed final one in a finite-time\cite{dech19,vu23a}. Likewise, the thermodynamic cost of information processing\cite{parr15} and of computation\cite{wolp24} can be cast in such a form. 
Whenever such an inequality involves coarse-grained observables, the framework outlined in this review should be applicable
to such a complementary perspective.

\subsection{Beyond the stochastic Markovian paradigm?}

The derivation of these inequalities -- be they used as bounds on entropy production or for the minimal cost of reaching a goal  -- relies crucially on the assumption that, on some deep level, the system obeys a stochastic Markovian dynamics even though on the observational level it may typically be non-Markovian, i.e., exhibit memory. One may wonder whether there exist further frameworks on which one can build a consistent thermodynamic description of 
driven classical systems.

Indeed, by alternatively assuming a Hamiltonian dynamics with driving implemented through a time-dependent potential, the Jarzynski relation yields a refinement of the second law in the Kelvin-Thomson version that relates applied work to a difference in free energy\cite{jarz11}.
The work is operationally accessible if all degrees of freedom to which the time-dependent driving is applied are
observable. An identification of thermodynamic entropy production in such a Hamiltonian approach, however, requires the further crucial assumption that correlations between system and bath are broken through a time-scale separation \cite{seif16}. 

Beyond these two settings -- a time-dependent Hamiltonian dynamics and an underlying stochastic Markovian dynamics -- it remains an open challenge to explore whether for classical systems further frameworks can be conceived from which a consistent comprehensive thermodynamics on a coarser scale could be derived. 
In view of the rapid methodological progress discussed in this review, however, the potential offered by the assumption of an underlying stochastic Markovian dynamics appears far from exhausted -- both in advancing theoretical understanding and in applying these concepts to the analysis of experimental data.

\acknowledgments
My research in this field has benefited from stimulating interactions with A.C. Barato, P. Pietzonka, T. Koyuk, J. Deg\"unther, B. Ertel, J. van der Meer, J.H. Fritz, and with A.M. Maier whom  I thank for a critical reading of this manuscript and for help with the figures.  
\newpage
%
 
\newpage

\newpage



\newpage~\newpage

\begin{shaded}

	{\bf{BOX 1: Stochastic thermodynamics for discrete states in a nutshell}}\cite{seif25}\\~\\
	
	On a microscopic scale, a system with degrees of freedom $\xib$ has an energy, i.e., Hamiltonian $H(\xib,\l)$, which later may become time-dependent through an external control parameter $\l =\l(t)$ denoting, e.g., the position of an optical tweezer. This system is embedded in, or in contact with, a heat bath of well-defined temperature $\TT$ with inverse $\beta\equiv 1/\kbT$,
	where Boltzmann's constant $\kb$ will be set to 1 throughout. Since microscopic states and their fast dynamics are neither observable nor typically informative, the focus on longer timescales shifts to a set of meso-states $\{i\}$. Formally, introducing these meso-states corresponds to an initial coarse-graining that can be described by a many-to-one 
	mapping $\xib \mapsto i$, for simplicity assumed to be independent of $\l$. At fixed $\l$, the system reaches equilibrium. The probability to find the system in meso-state $i$ then is 
	\beq
	p_i^e=\exp\{-\beta[F_i(\beta,\l)-F(\beta,\l)]\} .
	\ee
	Formally, the free energy $F_i(\beta,\l)$ of  meso-state $i$ is given by
	\beq
	F_i(\beta,\l) = -(1/\beta)\sum_{\xib\in i}\exp\{-\beta[H(\xib,\l)-F(\beta,\l)], \ee
	where $F(\beta,\l)=-(1/\beta)\sum_\xib \exp\{-\beta[H(\xib,\l)]\}$ is the free energy of this small system. Crucially, even without any knowledge of $H(\xib,\l)$, free energy differences $\Delta_{ij}F\equiv F_j-F_i$ are operationally accessible through the ratio of dwell times in these states taken from a  long equilibrium trajectory. Similarly, one can identify an intrinsic entropy $S_i$ and an internal energy $U_i$ for these meso-states through
	\beq S_i=\beta^2\partial_\beta F_i~~~\textrm{and} ~~~
	U_i=F_i+S_i/\beta.
	\label{eq:b3}
	\ee
	
	In biochemical systems, the meso-states may additionally involve molecules of species $\alpha$  bound or released from a reservoir with chemical potential $\mu_\alpha$.   
	Energy conservation of the total system comprising the system proper, the surrounding heat bath, and the chemical reservoirs entails
	the relation
	\beq Q_{ij}= U_i-U_j - \Delta_{ij} U\chem\res\ee
	for the heat $Q_{ij}$ dissipated  in a transition from $i$ to $j$.  The last term is the change in internal energy of the chemical reservoirs due to the bound or released molecules.
	
	An incremental change of the external control parameter $\l$  implies the input of mechanical work $\dbar W\mech=\partial_\l F_{i(t)} d\l
	$ when the system is currently in state $i(t)$. The corresponding differential first law $\dbar W\mech=dU_{i(t)}+ \dbar Q$ with (\ref{eq:b3}) then implies $\dbar Q=-\partial_\l S_{i(t)} d\l/\beta$.
	
	The crucial assumption of a time-scale separation between the microscopic degrees of freedom contributing to one meso-state and that of transitions between the meso-states  implies a Markovian, i.e., memory-less dynamics on their level. Moreover, the rate $k_{ij}$ for a transition from state $i$ to state $j$  becomes $\l$-dependent. For thermodynamic consistency, which means that under equilibrium conditions any initial distribution relaxes into thermal equilibrium,
	the rates must obey local detailed balance, i.e., 
	\begin{eqnarray}
		k_{ij}/k_{ji} &=& \exp(-\beta\Delta_{ij} F\tot)\nn\\
		&=& \exp[\beta(F_i-F_j+\sum_\alpha d^\alpha_{ij} \mu _\alpha - fd_{ij})] .
	\end{eqnarray} This ratio involves the total change of the free energy, which, in this case, comprises the free energy of both the system and the chemical reservoirs that may loose molecules ($d^\alpha_{ij}>0)$ or gain them 
	($d^\alpha_{ij}>0)$  at their respective chemical potentials. Furthermore, for molecular motors, the transition may lead to a step of length $d_{ij}$ against an externally imposed force $f$. 
	The dynamics on the meso-level then follows the master equation
	\beq
	\partial_tp_i(t)=\sum_j[-k_{ij}p_i(t)+k_{ji}p_j(t)].
	\ee 
	
	The total entropy change consists of the contributions from the heat bath, the chemical reservoir, the system’s intrinsic entropy, and the stochastic entropy.
	 The latter reads in differential form
	\beq ds = d[-\ln p_i(t)]_{|i=i(t)} ,
	\ee where $p_i(t)$ solves the master equation for the given initial distribution. 
	Averaging leads to  the mean instantaneous rate of total
	entropy production that can be written as
	\beq
	\sigma(t) = 
	\sum_{ij}p_i(t)k_{ij}(\l)
	\ln\frac{p_i(t)k_{ij}(\l)}{p_j(t)k_{ji}(\l)}\geq 0.
	\label{eq:sigma-l}
	\ee
	
	Finally, in a NESS, the path weight of a  trajectory $\gamma=[i(t)]$ starting at $i^0$ reads
	\beq
	p[\gamma|i^0]=\exp\left[-\sum_{ij}\tau_i k_{ij}\right]\prod_{ij} k_{ij}^{n_{ij}}.
	\ee This path weight depends on the accumulated dwell time $\tau_i[i(t)]$ the trajectory $\gamma$ spends in a state $i$ and on the number $n_{ij}[i(t)]$ of transitions from $i$ to $j$ it undergoes.

\end{shaded}

\newpage

\begin{shaded}

	{\bf{BOX 2: Markovian events and Markovian snippets: Generalizations of the waiting-time-based bound}}\\~\\
	A fine-grained trajectory $\gamma=[i(t)]$ or $\gamma=[\xb(t)]$ 
	follows a Markovian dynamics, meaning that at any given time the current state fully determines the probability of future evolution. In contrast, a coarse-grained trajectory $\Gamma$ typically exhibits memory effects: the probability of future behavior depends not only on the present state but also on its past. However, there can be specific moments along such a trajectory at which the future becomes independent of the past -- these are called Markovian events.
	
	One example of a Markovian event is the observation of a transition $I=i\to j$. Likewise, if the coarse-graining process leaves a fine-grained meso-state $i$ unaffected, then both exiting $i$ 
	into a lumped coarse-grained state and entering 
	$i$ from such a lumped state are also Markovian events. We will denote Markovian events with Greek indices $\alpha,\beta,\gamma,...$ and their time-reversed partners with a tilde.

	The parts of the coarse-grained trajectory between two consecutive of these events are called Markovian snippets.
	Further observations $\Oc_{\alpha\beta}$ along the snippet between the Markovian events $\alpha$ and $\beta$  may not qualify as Markovian since these observations do not determine the future independently of the past. 
	Such further data comprise, e.g.,  the visit of lumped meso-states. For a continuous dynamics, they may contain a visit of a region in phase space and the crossing of a manifold therein, see Fig. 2. These data then enter more detailed waiting-time distributions of the form 
	$\psi_{\alpha\beta}(\Oc_{\alpha\beta},t)$
	with normalization 
	\beq\sum_\beta\sum_{\Oc_{\alpha\beta}}\int_0^\infty dt \psi_{\alpha\beta}(\Oc_{\alpha\beta},t)
	=
	\sum_\beta p_{\alpha\beta} = 1.
	\ee 
	If these  observations have a unique behavior under time reversal, $\Oc_{\alpha\beta}\mapsto\tilde\Oc_{\tilde \beta\tilde \alpha}$, they can be used to improve the bound (\ref{eq:sig-wtd}) as\cite{vdm22b}
	\beq\sigma\wtd = \sum_{\alpha\beta}\sum_{\Oc_{\alpha\beta}}\int_0^\infty dt\nu_\alpha\psi_{\alpha\beta}(\Oc_{\alpha\beta},t)\ln\frac{\psi_{\alpha\beta}(\Oc_{\alpha\beta},t)}{\psi_{\tilde \beta \tilde \alpha}(\tilde \Oc_{\tilde \beta\tilde \alpha}, t)} .
	\label{eq:sig-wtd-o}
	\ee 
	
	A further step of coarse-graining can involve a many-to-one mapping of Markovian events $\{\alpha,\beta,\gamma,...\}$ into a smaller set of those denoted by $\{A,B,C,...\}$ as illustrated in Fig. 3. If this coarse-graining commutes with time reversal, one can derive the bound\cite{erte24}
\beq
\sigma\geq\frac{1}{2}
\left[\sum_{AB}\int_0^\infty dt\nu_A\psi_{AB}(t)\ln\frac{\psi_{AB}(t)}{\psi_{\tilde B\tilde A}(t)} +\sum_A \nu_A\ln\frac{\nu_A}{\nu_{\tilde A}}\right]
\label{eq:sig-wtd-be}
\ee that involves the rate $\nu_A$ with which events from set $A$ occur and the waiting-time distribution $\psi_{AB}(t)$ between events from two such sets.  Specific examples including chemical reactions networks are given in\cite{erte24,haru24}. 
\end{shaded}
\newpage
\begin{shaded}
	{\bf{BOX 3: Underdamped dynamics}}\\~\\
	In systems where inertia plays a role, the identification of physical entropy production is somewhat subtle\cite{spin12,lee12,fisc20,lyu24}. For pointing  out the salient issue, it will be sufficient to consider the one-dimensional case. A particle with mass $m$ and friction coefficient $\eta$ is driven by a non-conservative force $f$
	along a periodic potential $V(x,\l)$, which may be time-dependent through a control parameter $\l=\l(t)$. 
	The underdamped Langevin equation reads
	\beq
	m\dot v + \eta v = f -\partial_x V(x,\l) + \xi 
	\ee
	with $\dot x=v$ and 
	noise correlations
	\beq\langle \xi(t)\xi(t')\rangle = 2 (\eta/\beta)  \delta(t-t') .
	\ee
	
	The differential first law reads
	\begin{eqnarray}
		\dbar Q &=& \dbar W - dU\nn\\
		&=&[fdx +\partial_\l V(x,\l)d\l]- d[mv^2/2+V(x,\l)] \nn\\
		&=&F(x,\l)dx-mv dv
	\end{eqnarray}
	with total force $F(x,\l)=f-\partial_xV(x,\l)$.
	Likewise, the differential of stochastic entropy reads
	\beq
	ds = -d[\ln p(x,v,t)]_{|x=x(t),v=v(t)},\ee
	where $p(x,v,t)$ solves the Klein-Kramers equation
	\beq
	\partial_t p(x,v,t)=-\partial_x vp(x,v,t) -\partial_vj_v(x,v,t)
	\ee
	with the velocity-current
	\beq j_v(x,v,t)\equiv (1/m)[-\eta v+F(x,\l)+ (\eta/\beta m)\partial_v]p(x,v,t) 
	.\ee
	
	Integrated along a trajectory $\gamma\equiv [x(t),v(t)]$ with time-reversed counterpart
	$\tilde \gamma =[x(T-t),-v(T-t)]$,
	the total entropy production becomes
	\begin{eqnarray}
		\Delta s\tot[\gamma] &=& \beta Q + s|^t_0
		\nn\\	 &=&	
		\ln \{p[\gamma|\gamma^0]/\tilde p[\tilde \gamma|\tilde \gamma^0]\} + \ln [p (\gamma^0)/p(\gamma^T)]\nn\\
		&=&
		\ln \{p[\gamma]/\tilde p[\tilde \gamma]\} + \ln [p(\widetilde {\gamma^T})/p(\gamma^T)].
		\label{eq:box3-7}	
	\end{eqnarray}
	The distributions in the last term are the one at the final time, i.e., $p(\gamma^T)=p(x^T,v^T,T)$ and 
	$p(\widetilde {\gamma^T})=p(x^T,-v^T,T)$ and, note, $\tilde \gamma^0=\widetilde{\gamma^T}$.
	The additional contribution compared to (\ref{eq:sig-info}) arises from the odd character of the velocity. It is crucial for the correct identification of entropy production in finite time and implies, e.g., that the DFT $p(-\Delta s\tot)= p(\Delta s\tot)\exp(-\Delta s\tot)$ does not hold for  finite $T$, in general.
\end{shaded}

\newpage


\onecolumngrid 	
	
	\begin{figure}
		\centering
		\includegraphics[width=0.9
		\textwidth]{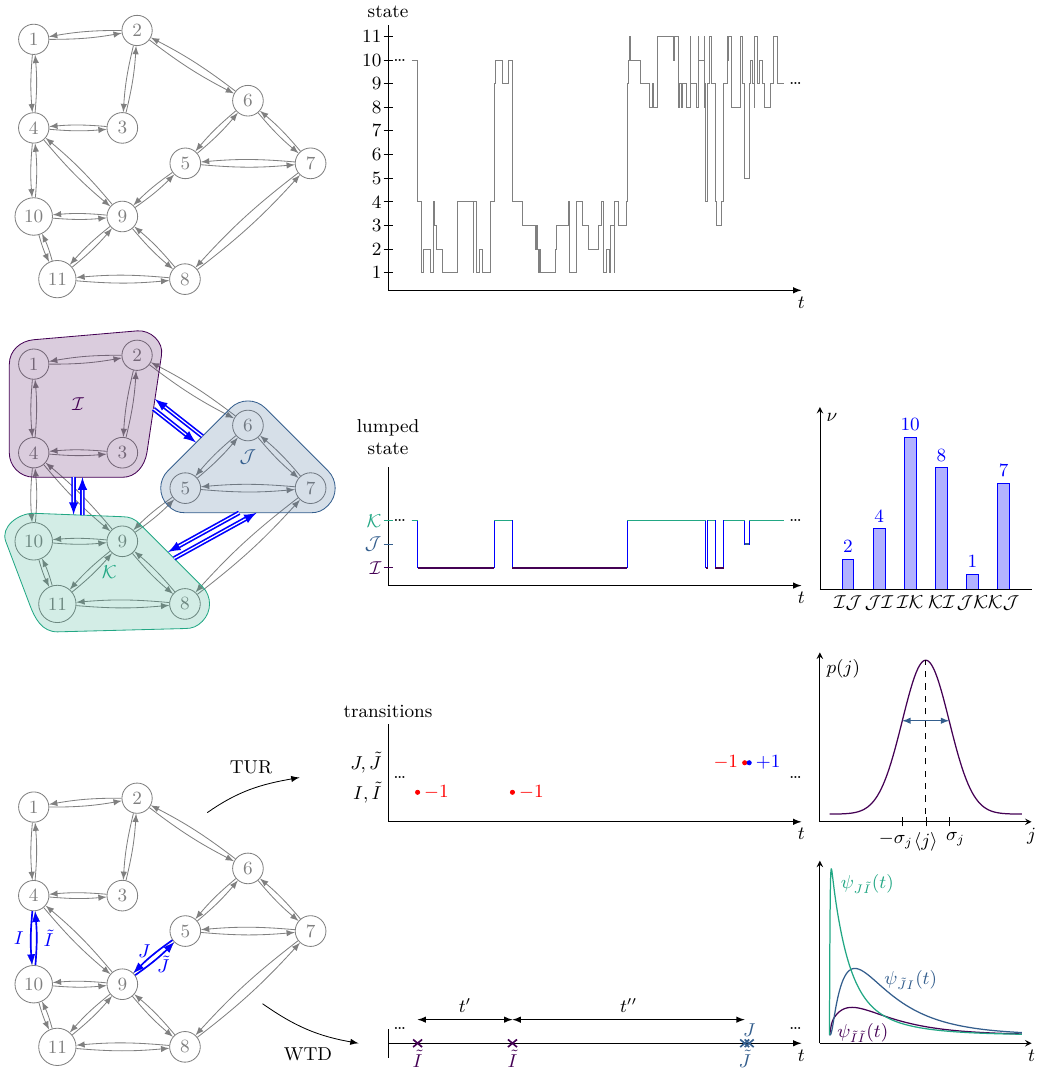}
		\caption{Several strategies for data extraction from coarse-grained trajectories based on partial observations. Top: An underlying mesoscopic network of 11 states and corresponding trajectory $\gamma=[i(t)]$.
			Middle: When only the three lumped states $\Ic,\Jc,\Kc$ are observed, the coarse-grained trajectory $\Gamma=[\Ic (t)]$ yields fluxes $\nu_{\Ic\Jc}$ (blue double arrows) between these three states that enter the lower bound $\sigma\app$  in (\ref{eq:sigma-app}). Bottom: This observer can record only  the transitions between states 4 and 10, denoted as $I$ and $\tilde I$,  and those between 5 and 6, denoted as $J$ and $\tilde J$.  Expressed as a time series of + and -- events, the mean and variance of the corresponding current $j$  (\ref{eq:j}), with $d_{4,10}=-d_{10,4}=d_{5,9}=-d_{9,5}=1$, enters the TUR-bound in $\sigma\tur$  (\ref{eq:tur}).
			For evaluating the bound $\sigma\wtd$ in (\ref{eq:sig-wtd}), 
			the waiting-time distributions $\psi_{IJ}(t)$ of two consecutive transitions are required. As an example, the time-intervals indicated by $t'$ and $t''$ contribute to $\psi_{\tilde I\tilde I}(t)$ and $\psi_{\tilde I\tilde J}(t)$, respectively.}
		\label{fig:one}
	\end{figure}
	
	\newpage	
	\twocolumngrid




\newpage

\newpage

\onecolumngrid

\begin{figure}
	\centering
	\includegraphics[width=0.5\textwidth]{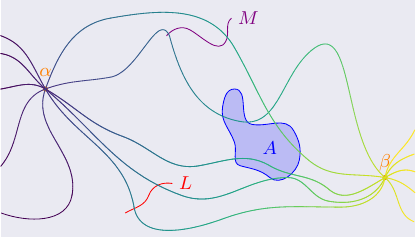}
	\caption{Markovian events, snippets  and additional data for a continuous dynamics. Several fine-grained trajectories pass through the point $\alpha$ and, after the interval $t$, through the point $\beta$ which denotes two Markovian events. Some of these trajectories additionally cross the manifolds $M$ or $L$, or visit the region $A$, which enter as data $\Oc_{\alpha\beta}$ the lower bound $\sigma\wtd$ given in (\ref{eq:sig-wtd-o}). For a coarse-grained trajectory $\Gamma$, the part between $\alpha$ and $\beta$ makes up one snippet.}
	
\end{figure}

\begin{figure}
	\centering
	\includegraphics[width=0.4\textwidth]{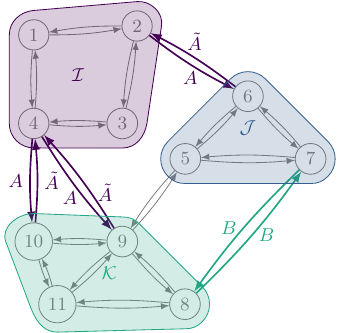}
	\caption{Coarse-graining Markovian events.
	The observer records
	transitions  from the set $A$ that leave the lumped state $\Ic$, and transitions from the set $\tilde A$ that  enters this state, without resolving the underlying fine-grained events contributing to $A=\{4\to 10, 4\to 9, 2\to 6\}$ and $\tilde A=\{10\to 4,9\to 4, 6\to 2\}$. 
	The observer also detects transitions between states 7 and 8 but cannot determine their direction, implying 
	$B=\tilde B=\{7\to 8, 8 \to 7\}$.
	The non-vanishing contributions to the first sum in (\ref{eq:sig-wtd-be}) are
	the pairs $AB$ and $B\tilde A$. The second sum runs over $A$ and $\tilde A$ for this example.}
\	
\end{figure}

\end{document}